\documentclass[twocolumn,prl,sort&compress,floatfix]{revtex4-1}
\usepackage{amssymb}
\usepackage{amsmath}
\usepackage{mathrsfs}
\usepackage{enumitem}
\usepackage{cleveref}
\usepackage{bm}
\usepackage{tabularx}
\usepackage[dvipsnames]{xcolor}
\usepackage[sort&compress]{natbib}
\usepackage{graphicx}
\usepackage{color}
\definecolor{red}{rgb}{1,0,0}
\definecolor{blue}{rgb}{0,0,1}
\usepackage{xr}
\DeclareMathOperator{\sgn}{sgn}
\renewcommand{\figurename}{\textbf{Figure}}

\newcommand{\red}[1]{\textcolor{black}{#1}}

\begin{document}

\title{Shaping Epigenetic Memory via Genomic Bookmarking: Supplementary Information}

\author{D. Michieletto$^{1,\dagger,*}$}
\author{M. Chiang$^{1,*}$}
\author{D. Col\`i$^{2,*}$}
\author{A. Papantonis$^{3}$}
\author{E. Orlandini$^{2}$}
\author{P. R. Cook$^{4}$}
\author{D. Marenduzzo$^{1,\dagger}$}
\affiliation{$^1$ School of Physics and Astronomy, University of 
	Edinburgh, Peter Guthrie Tait Road, Edinburgh, EH9 3FD, UK.\\ $^2$ Dipartimento di Fisica e Astronomia and Sezione INFN, Universit\'a di Padova, Via Marzolo 8, Padova 35131, Italy\\ $^3$ Centre for Molecular Medicine, University of Cologne, Robert-Koch-Str. 21, D-50931, Cologne, DE\\
 $^4$ The Sir William Dunn School of Pathology, South Parks Road, Oxford OX1 3RE, UK.\\
	$*$ Equal contribution; 
	$\dagger$ corresponding author}
	
\begin{abstract}
\textbf{Reconciling the stability of epigenetic patterns with the rapid turnover of histone modifications and their adaptability to external stimuli is an outstanding challenge. Here, we propose a new biophysical mechanism that can establish and maintain robust yet plastic epigenetic domains via genomic bookmarking (GBM). We model chromatin as a recolourable polymer whose segments bear non-permanent histone marks (or colours) which can be modified by ``writer'' proteins. The three-dimensional chromatin organisation is mediated by protein bridges, or ``readers'', such as Polycomb Repressive Complexes and Transcription Factors. The coupling between readers and writers drives spreading of biochemical marks and sustains the memory of local chromatin states across replication and mitosis. In contrast, GBM-targeted perturbations destabilise the epigenetic patterns. Strikingly, we demonstrate that GBM alone can explain the full distribution of Polycomb marks in a whole Drosophila chromosome. We finally suggest that our model provides a starting point for an understanding of the biophysics of cellular differentiation and reprogramming.}
\end{abstract}

\maketitle

\setcounter{section}{0}
\setcounter{figure}{0}
\setcounter{table}{0}
\setcounter{equation}{0}

\renewcommand{\figurename}{Fig. S}
\renewcommand{\tablename}{Table S}
\newcommand{\figref}[1]{Fig. S\ref{#1}}
\newcommand{\tableref}[1]{Table S\ref{#1}}

 \newcommand{\todo}{\textcolor{red}{TODO}}
 
\renewcommand{\vec}[1]{\bm{#1}}
\newcommand{\set}[1]{\left\{ #1_i \right\}_{i=1\dots M}}
\newcommand{\shortset}[1]{ #1_{i=1\dots M}}
\newcommand{\kratky}  {\ensuremath{\mathrm{U}_{\text{KP}}}}
\newcommand{\harmonic}{\ensuremath{\mathrm{U}_{\text{H}}}}
\newcommand{\heaviside}[1]{\Theta\left(#1\right)}
\newcommand{\ULJ}     {\ensuremath{\mathrm{U}_{\text{LJ}}^{\text{tot}}}}
\newcommand{\norm}[1]{\left\|#1\right\|}
\newcommand{\abs}[1]{\left|#1\right|}

\section{Introduction}
%intro
%%motivation
%Transcriptional regulation underlies the ability of genomes to encode for different states associated with the same DNA sequence. For instance, 
Cells belonging to distinct tissues in a multi-cellular organism possess exactly the same genome, yet the DNA sequence is expressed differently. This is made possible by the establishment of lineage-specific epigenetic patterns (or ``landscapes'') -- the heritable marking of post-translational modifications (PTM) on histones and of methylation on DNA~\cite{Waddington1942,Alberts2014,Strahl2000,Kakutani2001,Cavalli2013, Pal2013,Probst2009,Ng2008}.  
%inheritable but plastic
%molecular basis of epigenetics
%%Histone modifications are dynamic 
Epigenetic patterns are robust, as they can be remembered across many rounds of cell division~\cite{Alberts2014,Waddington1942,Angel2011,Waddington1942, Kouskouti2005a,Ciabrelli2017,Probst2009}. 
At the same time, they are plastic and dynamic. They can adapt in response to external stimuli~\cite{Waddington1942,Stern2012,Wood2014,Angel2011,Feuerborn2015}, and they are affected by disease and ageing~\cite{Pal2016,Heard2014}. Additionally, many biochemical marks encoding the epigenetic information can turn over rapidly and are lost during DNA replication~\cite{Zentner2013,Klosin2017}. For example, acetyl groups on histones have half-lives $<10$ minutes~\cite{Zentner2013,Barth2010}, methyl groups on histones change during the period of one cell cycle~\cite{Kheir2010,Alabert2015,Zentner2013} and DNA methylation is modified during development~\cite{Heard2014}. The turnover may originate from histone replacement/displacement during transcription~\cite{Zentner2013,Skene2013,Probst2009,Festuccia2017}, replication~\cite{Scharf2009,Klosin2017,Probst2009} or from stochastic PTM deposition and removal~\cite{Arnold2013,Dodd2007,Berry2017}. 
%Waddington=heat shock genes in drosophila
%wood2014=seasonal sheep mating
%Angel2011=vernalisation
%ciabrelli2017=cavalli paper on drosophila
%coleman h3k27 memory -- science
% wang h3k9 memory -- science
% laprell hk27 memory eviction of pre -- science

%%FIG 1 Model 
\begin{figure*}[t!]
	\centering
	\includegraphics[width=0.9\textwidth]{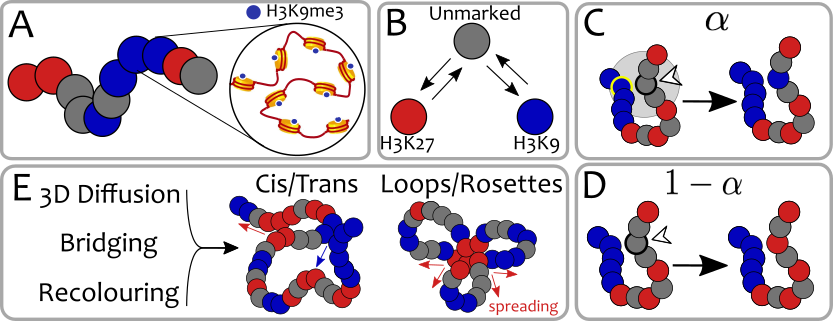}
	\caption{\textbf{Polymer Model with Dynamic Epigenetic Patterns.} \textbf{(A)} In our coarse-grain polymer model, each bead represents a group of nucleosomes and its colour captures the predominant epigenetic mark. \textbf{(B)} Epigenetic marks are dynamic. They can change between red, blue or grey (no mark) according to biophysical rules. For example, a red bead can be thought of as an inactive Polycomb state (marked by H3K27me3) while a blue bead as a heterochromatic segment (marked by H3K9me3). The precise nature of the marks does not affect the qualitative behaviour of this generic model. In the Voter-like dynamics, each bead must go through the unmarked state (grey) before changing to the opposite colour~\cite{Dodd2007}. Each bead is selected at rate $k_R$ (see text and SM) and, \textbf{(C)} with probability $\alpha$, it changes its colour ``closer'' to that of a randomly chosen 3D-proximal bead (in this case the one circled in yellow, see also SM). \textbf{(D)} The same bead has probability $1-\alpha$ to undergo a random colour conversion (in this case to red, see SM). \textbf{(E)} The synergy between 3D chromatin dynamics, bridging due to (implicit) binding-proteins/TFs and epigenetic recolouring gives rise to dynamic structures such as loop/rosettes and cis/trans contacts which drive (cis and trans) epigenetic spreading (indicated by red/blue arrows, see text). }
	\label{fig:model}
\end{figure*}

%Epigenetic information is disseminated along the chromosomes in the form of biochemical marks deposited on histones~\cite{Alberts2014}. 
Our goal is to develop a biophysical model that can reconcile the reproducible and robust formation of heritable yet plastic epigenetic landscapes across cell populations in the face of the rapid turnover of the underlying histone marks. In particular we will be interested in models which can yield ``epigenetic domains'', by which we mean 1D stretches of similarly-marked histones which tend to be co-localised in 3D and co-regulated~\cite{Sexton2012,Dixon2012,Jost2014B,Rao2014,Dixon2015}. [Note that in the context of our model, the terms histone marks, chromatin states and PTM will be used interchangeably.]

%\newpage
%Problem with current models
Existing models describe changes of PTMs in one-dimension (1D) or through effective long-range contacts; they yield smooth transitions between stable states and weak (transient) bistability~\cite{Dodd2007,Micheelsen2010,Dodd2011,Anink-Groenen2014,Obersriebnig2016,Jost2014B,Arnold2013,Erdel2016,Erdel2013, Erdel2017}. In contrast, our model explicitly takes into account the realistic structure and dynamics of the chromatin fibre in 3D (Fig.~\ref{fig:model}) -- crucial elements for the spreading of histone marks \emph{in vivo} ~\cite{Talbert2006,Lanzuolo2007,Engreitz2013,Pinter2012,Schauer2017, Ciabrelli2017,Deng2014}.

%%KEEP THIS NOTE FOR POSSIBLE REPLY %%%
%Architectural proteins like CTCF have been postulated to act as epigenetic insulators; on the other hand, they have been proven effective only in the case of 1D models but fail to create stable epigenetic domains when 3D spreading is accounted for~\cite{Dodd2011,Engreitz2013}.
%%%%%%%%%%%%%%%%%%%%%%%%%%%%%%%%%%%%%%%%%%5

% **************************************************************
% Keep this command to avoid text of first page running into the
% first page footnotes
\enlargethispage{-65.1pt}
% **************************************************************

From the physical perspective, accounting for realistic 3D interactions (e.g., the formation of loops and trans-contacts driven by the binding of bi- and multi-valent transcription factors) triggers ``epigenetic memory''~\cite{Probst2009,Ng2008}, i.e., stability of the epigenetic patterns against extensive perturbations such as DNA replication~\cite{Michieletto2016prx}. Within this framework, the possible ``epigenetic phases'' of the system are either disordered (no macroscopic epigenetic domain is formed) or homogeneous (only one histone mark spreads over the whole chromosome). Thus, no existing biophysical model can currently predict the spontaneous emergence of multiple heritable epigenetic domains starting from a ``blank'' chromatin canvas~\cite{Michieletto2016prx}. 

%what we do -- quick description of model and assumptions
Here, we propose a model for the {\it de novo} formation, spreading and inheritance of epigenetic domains that relies solely on three elements. First, we assume a positive feedback between multivalent PTM-binding proteins (``readers'') and other proteins which replace such marks (``writers''). 
%Examples for the assumptions
This captures the well-known observations that, for instance, HP1 (a reader binding to heterochromatin) recruits SUV39h1 (a writer for H3K9me3~\cite{Hathaway2012}), and that the Polycomb-Repressive-Complex PRC2 (a reader) contains the enhancer-of-zeste EZH2 (a writer) that spreads H3K27me3~\cite{Angel2011,Zentner2013,Hauri2016,Collinson2016,Michieletto2016prx}. 
Second, we assume the presence of genomic bookmarking (GBM) factors, typically transcription factors that can bind to their cognate sites and remain dynamically associated with chromatin through mitosis~\cite{Teves2016}. Examples of such GBMs include Polycomb-Group-Proteins (PcG)~\cite{Kassis2013,Schuettengruber2014,Ciabrelli2017,Laprell2017}, and Posterior-Sex-Combs (PSC)~\cite{Follmer2012} bound to Polycomb-Response-Elements (PREs) in \emph{Drosophila}~\cite{Lanzuolo2007,Laprell2017,Follmer2012,Ciabrelli2017}, GATA~\cite{Kadauke2012,Kadauke2013a} and UBF~\cite{Grob2014} in humans and Esrbb~\cite{Festuccia2016,Festuccia2017} and Sox2~\cite{Teves2016,Deluz2016} in mouse. Here, we will use the term transcription factor (TF) to include both activators and repressors. 
Third, we assume that the recruitment of read-write machineries is coupled to specific GBM binding. 
These three assumptions allow our model to reconcile short-term turnover of PTM with long-term epigenetic memory and plasticity. Finally, we show that our model can quantitatively recapitulate the distribution of H3K27me3 mark seen in Drosophila cells \emph{in vivo}.

%%% KEEP 
%\footnote{It is widely assumed that most TFs dissociate from chromosomes during mitosis; however, this assumption turns out to be incorrect. Thus, live-cell imaging shows that many factors previously thought to be lost during mitosis (e.g., Sox2, Oct4, Klf4, and Foxo1/3a) actually remain bound. The apparent loss was traced to a fixation artifact – as paraformaldehyde enters cells, it removes factors from the soluble pool to bias exchange with bound ones, and this strips bound molecules from chromosomes~\cite{Teves2016}.}
%%%

%%Fig2 phase diagram + 3d snaps
\begin{figure*}[t!]
	\centering
	\includegraphics[width=0.9\textwidth]{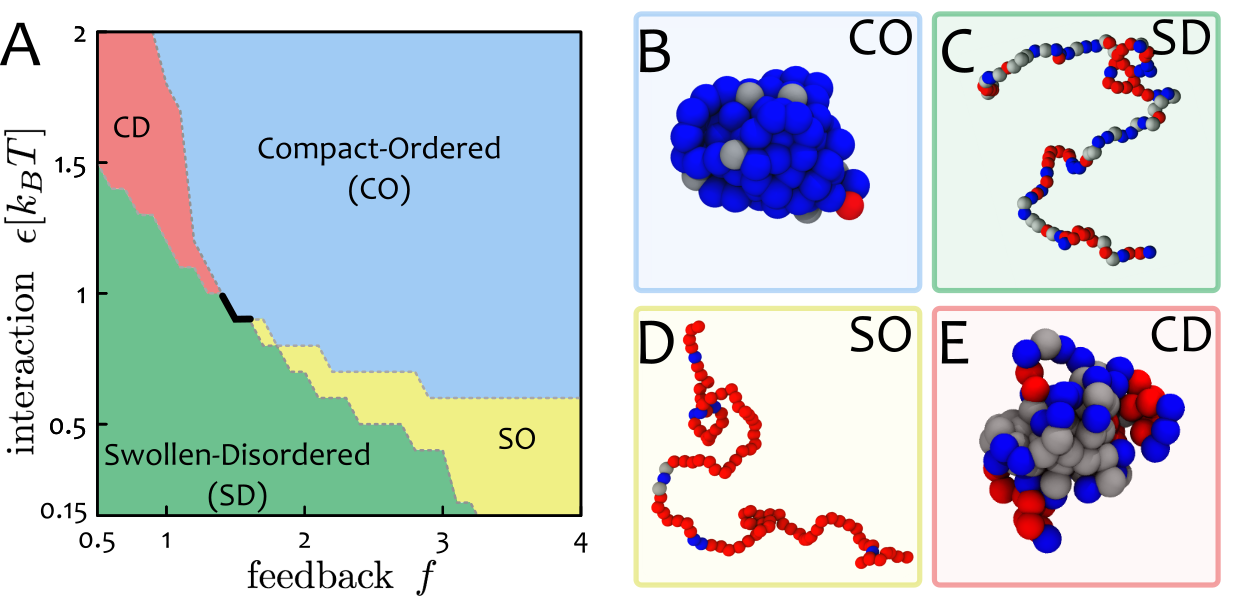}
	\caption{\textbf{Phase Diagram: Chromatin States and Epigenetic Memory.}  \textbf{(A)} The phase diagram of the system in the space $(\epsilon,f\equiv\alpha/(1-\alpha))$ displays four distinct regions: (i) swollen-disordered (SD); (ii) compact-ordered (CO); (iii) swollen-ordered (SO) and (iv) compact-disordered (CD). The thick solid line represents a first-order transition between the SD and CO phases, whereas the dashed lines signal smoother transitions between the regions.  (\textbf{B-E}) Representative snapshots of the stable states, which resemble conformations of chromatin seen \emph{in vivo}. The CO phase may be associated to globally-repressed heterochromatin, the SO phase to open transcriptionally-active euchromatin, and the the CD phase to ``gene deserts'' characterised by low signal of PTMs and collapsed 3D conformations~\cite{Sexton2012,Dixon2012,Kharchenko2011,Gilbert2006}.  The first-order nature of the SD-CO transition entails ``epigenetic memory''~\cite{Ng2008}, and the CO phase is robust against extensive perturbations such as the ones occurring during replication~\cite{Michieletto2016prx}.		
}
	%%%%%%%%%%%XXX write caption XXX
	%\textbf{(C)} Monte-Carlo model: a color change is weighted by the overall change in energy $\Delta U$ (see text). Top row: the bead indicated by the arrow is attempted to become blue; the move is accepted as $\Delta U < 0$. Bottom row: the bead indicated by the arrow is attempted to become red; since $\Delta U >0$ the colour change occurs with probability $p = \exp{(-\Delta U/k_BT)}$.   
	\label{fig:phasediagr}
\end{figure*}

\section{Material and Methods}

\subsection{A Polymer Model for Dynamic Epigenetic Patterns}

To capture the dynamic nature of epigenetic landscape due to PTM turnover and histone displacement~\cite{Festuccia2016,Zentner2013}, we enhance the (semi-flexible) bead-spring polymer model for chromatin~\cite{Rosa2008,Fudenberg2016,Mirny2011,Brackley2016nar, Brackley2016nucleus,Brackley2013pnas,Cheng2015,Sanborn2015,Rosa2010} by adding a further degree of freedom to each bead. Specifically, each bead -- corresponding to one or few nucleosomes \red{(choosing a different coarse-graining leaves our result qualitatively unaffected)} -- bears a ``colour'' representing the instantaneous local chromatin state (e.g., H3K9me3, H3K27me3, H3K27ac, etc., see Fig.~\ref{fig:model}(A)), which can dynamically change in time according to realistic biophysical rules~\cite{Michieletto2016prx,Dodd2007,Arnold2013} (see Fig.~\ref{fig:model}(B)). This is in contrast with previous works that only accounted for static epigenetic patterns via co-polymer modelling~\cite{Jost2014B,DiPierro2016,Brackley2016nar,Barbieri2012}.

We first consider a toy model in which beads may be found in one of three possible states: grey (unmarked), red (e.g., Polycomb-rich) and blue (e.g., heterochromatin-rich). [A more realistic model will be discussed later]. Beads bearing the same histone mark are mutually ``sticky'', indicating the presence of implicit bridging proteins~\cite{Zentner2013,Brackley2016nar,Brackley2013pnas}, and can thus bind to each other with interaction energy $\epsilon$ (see Fig.~\ref{fig:model}(E)). All other interactions are purely repulsive. 
\red{The natural time-scale for our simulations is the Brownian time $\tau_{Br}=\sigma^2/D$ which is the typical diffusion time for a bead of size $\sigma$. As discussed in the SM, this time can be estimated as $\tau_{Br}\simeq 10$ ms which is equivalent to considering a nucleoplasm viscosity of $\eta=150$ cP and a bead of size $\sigma=30$ nm~\cite{Michieletto2016prx}.}

The action of writer proteins is modelled through ``recolouring'' moves occurring at rate $k_R$; here, we set $k_R=0.1 s^{-1}$ which is close to typical timescales for acetylation marks~\cite{Barth2010}. In selected cases, we have also employed a faster recolouring rate of $k_R= 10 s^{-1}$ to ensure faster convergence to steady state (see SM for details on simulations and time-mapping). 

Our model couples reading and writing as follows. First, a bead is selected randomly. Next, with probability $\alpha$, it recruits a neighbour from spatially-proximate beads (within $r_c=2.5 \sigma$, where $\sigma$ is bead size). The colour of the first bead is then shifted one step ``closer'' to the colour of the second (Fig.~\ref{fig:model}B-C). Otherwise (with probability 1-$\alpha$), the bead undergoes a noisy conversion to one of the other colours (see Fig.~\ref{fig:model}D and SM for further details). 

%%%%MOVE TO SM %%%%%%%%%%%%%%%5
%In the first pathway (P1), a bead is selected randomly (with uniform probability), and a recolouring move proposed; the new energy of the system is computed and the move is accepted or rejected according to the Metropolis criterion (see Fig.~\ref{fig:model}(C) and Methods (SM) or Ref.~\cite{Michieletto2016prx}). There is an effective ``recolouring'' temperature $T_R$ associated with this move and another temperature ($T_L$, see Methods, which may in principle be different) controls the 3D chromatin dynamics. 

%Whereas pathway 1 may describe a process that is in thermodynamic equilibrium (if $T_L=T_R$),
%%%%%%%%%%%%%%%%%%%%%%%%%%%%%%%%%%

This re-colouring scheme encodes a purely non-equilibrium process and it is akin to a ``voter'' or ``infection-type'' model~\cite{Dodd2007,Arnold2013}. In SM, we describe a ``Potts'' recolouring scheme which can be arbitrarily tuned either in- or out-of-equilibrium~\cite{Michieletto2016prx}. Both schemes couple 1D epigenetic information along the chromatin strand to 3D folding. Both drive a positive feedback loop between readers (which bind and bridge chromatin segments) and writers (which can change the underlying epigenetic pattern). Strikingly, both strategies lead to qualitatively similar behaviours, in which cis/trans contacts, globules and rosettes (Fig.~\ref{fig:model}E) spontaneously emerge and drive the spreading of histone modifications. 
% thereby proving the robustness of our model
To simplify the presentation of our results, and because the observed behaviours are similar, we choose to report in the main text the finding obtained via the ``infection-type'' model. This model may better capture the one-to-one nature of the chemical reactions required for the deposition (or writing) of histone marks (see SM for more details). 
%The results obtained with the ``Potts'' recolouring scheme are presented in the SM. Explicit mention to the two models will be made where the two lead to different behaviours.

\section{Results}

\subsection{The Phase Diagram of the System Entails Epigenetic Memory}

We first map the phase diagram obtained by varying the ``feedback'' parameter $f=\alpha/(1-\alpha)$ and the attraction energy $\epsilon/k_BT$ between any two like-coloured beads. A more realistic model accounting for different attractions between ``Polycomb-rich'' and ``heterochromatin-rich'' beads is considered later.

Figure~\ref{fig:phasediagr}A shows that there are four distinct phases predicted by our minimal model. First, at small $\alpha$ and $\epsilon/k_BT$, the fibre is swollen and epigenetically disordered (SD). At large $\alpha$ and $\epsilon/k_BT$, the system is in the compact epigenetically ordered (CO) phase. These two states are separated by a discontinuous transition, signalled by the presence of hysteresis and coexistence (see SM). The discontinuous nature of the transition is important because it confers metastability to the two phases with respect to perturbations in the parameter space. In addition, perturbing a compact heterochromatin-rich state by extensively erasing PTM marks (e.g. during replication) fails to drive the system out of that epigenetic state~\cite{Michieletto2016prx}; in other words, the global epigenetic state is remembered across genome-wide re-organisation~\cite{Michieletto2016prx,Angel2011}.   

The two remaining regions of the phase diagram (Fig.~\ref{fig:phasediagr}A) are (i) an ordered-swollen phase (SO), observed at large $\alpha$ but small or moderate $\epsilon/k_BT$, and (ii) a compact-disordered phase (CD), found at small $\alpha$ and large $\epsilon/k_BT$. Our simulations suggest that the transitions from, or to, these states are smooth and unlike that between the SD and CO phases.

\red{We highlight that the first order line (black thick line in Fig.~\ref{fig:phasediagr}A) entails hysteresis (see SM, Fig.~S3) and robustness of the states against small perturbations in the parameter space. On the other hand, a pathway that brings, for instance, a CO state into a SD one passing through the SO region, crosses continuous lines. Such a pathway in the parameter space may be a valid model to describe a change of identity of a cell, for instance during reprogramming. While this is an appealing avenue, we leave its exploration for future work as it requires a more detailed mapping between the recolouring rules of real systems and our parameter space.}

%and either akin to a $\theta$-collapse of a flexible homopolymer~\cite{Grosberg1994} or to an Ising-like (second-order) transition on a 3D network of spins~\cite{Huang1987}.

%this stuff below is too technical/uninteresting for a biological audience
%The three transitions, between swollen ordered and compact ordered, between compact disordered and compact ordered, and between swollen disordered and compact disordered are all continuous. The first transition (swollen to compact ordered) is analogous to the $\theta$ collapse of a homopolymer, which is a smooth transition for flexible fibres as considered here. The second transition (compact disordered to order) is akin to a magnetic Ising-like transition in a 3D network of spins, which is known to be continuous~\footnote{Although this refers to the equilibrium Ising model only, our simulations suggest that our non-equilibrium model behaves similarly.}. Finally, the transition between swollen disordered and compact disordered, which is similar to the $\theta$ collapse in a random heteropolymer, also bears the signature of a second order transition in our simulations (no hysteresis, see SI and accompanying discussion). 

\subsection{Polymer Simulations of the Minimal Model Capture Realistic Chromatin Conformations}

Intriguingly, some of the phases in the phase diagram in Fig.~\ref{fig:phasediagr} correspond to structures seen in eukaryotic chromosomes. Most notably, the compact-ordered phase provides a primitive model for the structure of the inactive copy of the X chromosome in female mammals; this is almost entirely transcriptionally silent, and this state is inherited through many cell divisions~\cite{Alberts2014}.

The compact-disordered phase is reminiscent of ``gene deserts'' (or black chromatin~\cite{Sexton2012,Kharchenko2011}). \red{This is a state without a coherent epigenetic mark which tends to co-localise in 3D, possibly due to the linker histone H1~\cite{Sexton2012,Kharchenko2011,Filion2010}}.  
Finally, the swollen-ordered phase is reminiscent of open and transcriptionally-active chromatin~\cite{Gilbert2004,Gilbert2006,Nozawa2017}. 

In this simplified model, feedback between readers and writers leads to unlimited spreading of a single histone mark in both ordered phases (CO and SO, see Fig.~\ref{fig:phasediagr})~\cite{Michieletto2016prx,Michieletto2017scirep}. Although near-unlimited spreading of silencing marks is seen in telomere position effects in yeast~\cite{Talbert2006} and position-effect variegation in \emph{Drosophila}~\cite{Schotta2002}), this minimal model cannot recapitulate the existence of multiple epigenetic domains, or ``heterogeneous'' epigenetic patterns.

% Instead, boundaries between domains disappear, and, eventually a single domain takes over the genome~\cite{MichielettoSciRep2017,Michieletto2016prx}.

%%Fig 3 kymo + bookmarks --> patterns 
\begin{figure*}[t!]
	\centering
	\vspace*{0.2 cm}
	\includegraphics[width=1\textwidth]{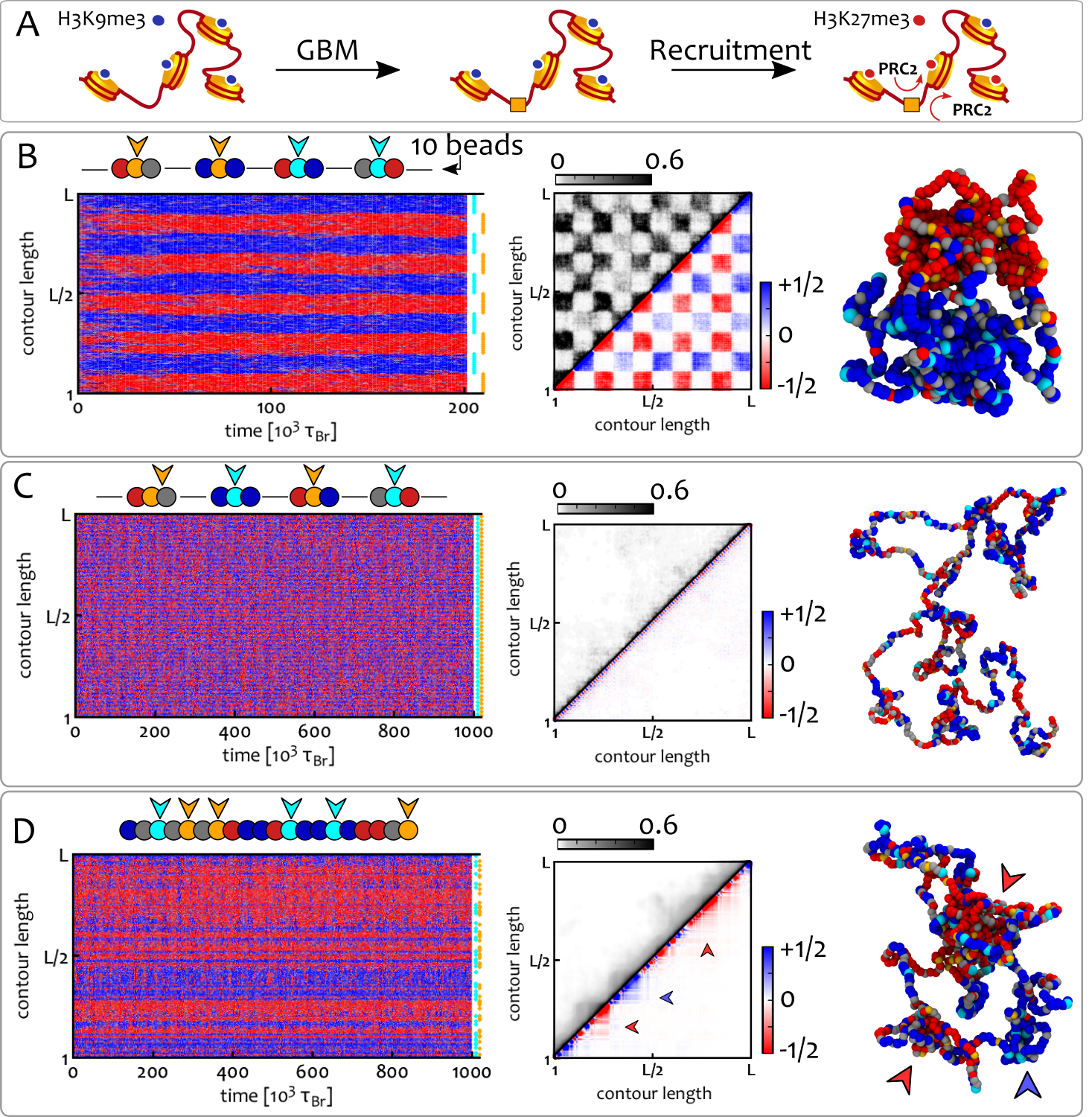}
	\vspace*{0.2 cm}
	\caption{\textbf{GBM Shapes the 1D Epigenetic Pattern and the 3D Chromatin Conformation.} (\textbf{A}) At the nucleosome level, GBM is mediated by a TF that binds to its cognate site and recruits read/write machineries that spread the respective histone mark to 3D-proximal histones (here PRC2 spreads H3K27me3). (\textbf{B-D}) We consider a chromatin fibre $L=1000$ beads long, starting from an epigenetically random and swollen condition with $\phi=0.1$, equivalent to one bookmark in 150 nucleosomes at 3kbp resolution and we fix $f=2$ and $\epsilon/k_BT=0.65$. GBM is modelled by imposing a permanent colour to some beads along the fibre. Cyan and orange beads denote bookmarks for blue and red marks, respectively. Plots show kymographs (left column), average contact maps (central column) and typical snapshots (right column) for different bookmarking patterns (shown at the end of kymographs and cartoons above). Contact maps are split into two: the upper triangle shows a standard heat-map quantifying the normalised frequency of contacts between segments $i$ and $j$, whereas the lower triangle shows an ``epigenetically-weighted'' one in which each contact is weighted by the type of beads involved (+1 for blue-blue contacts, -1 for red-red and 0 for mixed or grey-grey). (\textbf{B}) A clustered GBM pattern yields well-defined epigenetic domains which coalesce into A/B compartments ($k_R=0.1 s^{-1}$). (\textbf{C}) Alternate GBM maintains the chromosome in a swollen-disordered state ($k_R=10 s^{-1}$). (\textbf{D}) Random GBM creates stable and coexisting locally-compacted structures (indicated by the arrowheads) without generating long-range contacts ($k_R=10 s^{-1}$). See also Suppl. Movies 2-4 to appreciate the dynamics of the model.}
	\label{fig:bookmarks1}
\end{figure*}
 
\subsection{A Biophysical Model for Genomic Bookmarking}
We now introduce genomic bookmarking (GBM) to account for heterogeneous epigenetic patterns, coexistence of heritable epigenetic domains and active/inactive (A/B) compartments~\cite{Rao2014,Dixon2015}. A bookmark is here considered as a TF (activator or repressor) that binds to a cognate site and recruits appropriate readers or writers (see Fig.~\ref{fig:bookmarks1}A).

A mechanistic model of how bookmarks might guide the re-establishment of the previous epigenetic patterns after mitosis remains elusive~\cite{Heard2014,Kadauke2013a,Sarge2005,Teves2016}. Here, we assume that GBMs are expressed in a tissue-specific manner and remain (dynamically) associated to chromatin during mitosis~\cite{Teves2016,Follmer2012}. Then, on re-entering into inter-phase, they can recruit appropriate read/write machineries and re-set the previous transcriptional programme.

In our polymer model, we account for bookmarks by postulating that some of the beads cannot change their chromatin state (Fig.~\ref{fig:bookmarks1}A). Thus, a red (blue) bookmark is a red (blue) bead that cannot change its colour, and otherwise behaves like other red (blue) beads. In Figure~\ref{fig:bookmarks1}A, a bookmark is indicated by an orange square that binds to DNA (rather than a PTM) and recruits read/write machineries (e.g., PRC2), which then spread a histone mark (e.g., H3K27me3) to the neighbours~\cite{Alberts2014,Cheutin2012,Zentner2013,Cavalli2013}. 

It is important to stress that, in these polymer simulations, spreading of a colour is driven by the local increase in the density of that color. Indeed, bridging drives like-colour attractions and increases the probability that a random bead will be ``infected'' by a 3D-proximal bead bearing that mark. The choice of which mark dominates the local spreading is decided via symmetry breaking and we thus bias the local concentration of marks by introducing DNA-bound enzymes, i.e. bookmarks  (see Supplementary Movie 1).

%In turn, this leads to a re-establishment of the correct epigenetic landscape. 
% hence the presence/function of protein factors/higher-order conformations acting as physical barriers is justified to stratify such heterogeneity along the fibre

\subsection{GBM Drives Stable Coexistence of 1D Epigenetic Domains and Shapes the 3D Chromatin Organisation}

We now consider a chromatin fibre where a fraction $\phi$ of beads are ``bookmarks'' and analyse how their spatial distribution affects the epigenetic patterns in steady state. We consider three possible GBM distributions as follows: (i) {\it Clustered}: bookmarks are equally spaced along the fibre; the colour alternates after every $n_c$ consecutive bookmarks ($n_c>1$ defines the cluster size). (ii) {\it Mixed:} same as clustered, but now colours alternate every other bookmark ($n_c=1$). (iii) {\it Random:} random bookmarks are placed along the fibre while the fraction $\phi$ is kept constant. 

Figures~\ref{fig:bookmarks1}B-D show the results for  $\phi=0.1$ and a chromatin fibre $L=1000$ beads long. This correspond to about $3$ Mbp, or $1.5 \times 10^4$ nucleosomes, for a coarse graining of $3$ kbp per bead, i.e., a fibre with approximately one bookmark every $150$ nucleosomes. Simulations are initialised with the chromatin fibre in the swollen-disordered phase and non-bookmarked regions contain equal numbers of red, blue and grey beads. 

The clustered distribution of bookmarks (Fig.~\ref{fig:bookmarks1}B) reaches a stable epigenetic pattern with blocks of alternating colours (domains). On the contrary, the mixed bookmark distribution hinders domain formation, and the fibre remains in the SD state (Fig.~\ref{fig:bookmarks1}C). Remarkably, random bookmarks also yield domains in 1D (Fig.~\ref{fig:bookmarks1}D), even in the absence of any correlation between the location of bookmarks. 

%For both clustered and randomly scattered bookmarks, domain formation is fast, and occurs within $10^3 \tau_{\rm Br}$. 
%Stable domains cannot be formed throughout all parameter space (see SM): if $\epsilon/k_BT$ is too large, 3D spreading of red and blue marks leads to an ``all-or-none'' scenario (as in Fig.~\ref{fig:phasediagr}) in which a single epigenetic state takes over in steady state. Intriguingly, epigenetic domains are most stable in the neighbourhood of the transition between the swollen disordered and compact ordered phases. It is here that the 1D pattern of epigenetic domains on the fibre can be most reproducibly established and maintained dynamically simply by laying down a suitable pattern of bookmarks.

Importantly, we highlight that the bookmarking pattern affects 3D structure. Thus, in Figure~\ref{fig:bookmarks1}C-D, both the random and mixed patterns yield swollen or partially-collapsed fibres, even though the parameters used normally drive the system to a collapsed phase. [Note that our parameter choice accounts for the fact that the critical $\epsilon(f)$ marking the SD-CO transition decreases with $L$.] 

\red{For the random distribution, the contact map exhibits locally compact structures with coherent epigenetic marks (see arrowhead in Fig.~\ref{fig:bookmarks1}D) while long-range interactions between like-coloured domains are supressed}. This result is in marked contrast with equilibrium models with static epigenetic pattern~\cite{Jost2014B,Barbieri2012}). On the other hand, for clustered bookmarks, red and blue domains separately coalesce in 3D (macro-phase-separation), to give a checker-board appearance of the contact map (Fig.~\ref{fig:bookmarks1}B) reminiscent of the pattern formaed by A/B compartments in Hi-C maps after suitable normalisation~\cite{Lieberman-Aiden2009,Rao2014}. 

\red{We highlight that these patterns are achieved independently of the chosen initial configuration. As shown in the SM (Fig.~S4), a system initialised from deep into the collapsed-disordered phase (reminiscent of condensed mitotic chromosomes) leads to the same 1D pattern of marks and 3D organisation found in Fig.~\ref{fig:bookmarks1} at large times.}

\vspace{2 cm}
%We thus suggest that the epigenetic landscape in eukaryotic cells may then reflect a subtle combination of bookmarking patterns to give local TAD structures embedded within a genome which is globally (micro-)phase-separated into A/B-compartments.

\newpage
 \subsection{A Critical Density of Bookmarks is Required to Form Stable Domains}

\begin{figure}[t!]
	\centering
	\includegraphics[width=0.5\textwidth]{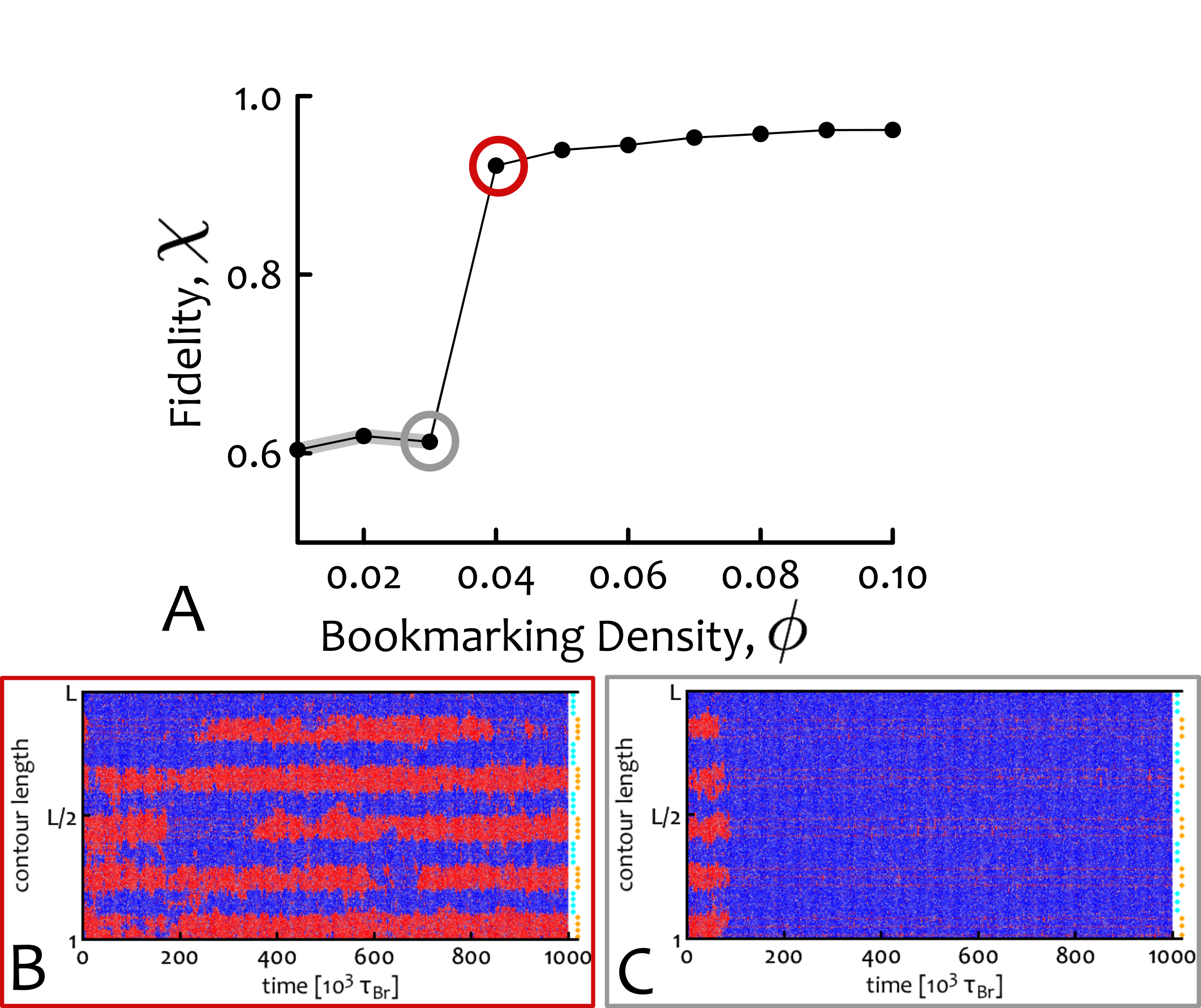}
	\caption{\textbf{A Critical Density of Bookmarks is Required for Stable Domain Formation.} (\textbf{A}) Using the clustered pattern of bookmarks at different densities $\phi$, we quantify the deviation from a ``perfect'' block-like epigenetic pattern. To do this we define the ``fidelity'', $\chi$, as $1-\Delta^2$ where $\Delta^2=Var \left[P_{\rm red}(i),\Pi(i)\right]$, i.e. the variance of the probability $P_{\rm red}(i)$ of observing a red bead at position $i$ with respect to the perfect square wave $\Pi(i)=0.5\left[\sgn\left(\sin{\left(\pi  i/n_d\right)}\right)+1\right]$, where $n_d$ is the number of beads in a domain (here $n_d = 100$). The fidelity $\chi$ jumps abruptly from a value near its lower bound of $1/2$ towards unity, at the critical $\phi_c\simeq 0.04$. (\textbf{B,C}) Kymographs representing the behaviour of the system at the points circled in red and grey in (\textbf{A}).  }
	\label{fig:accuracy}
\end{figure}

%The results reported in Figure~\ref{fig:bookmarks1} demonstrate that domain formation is only observed when there are many bookmarks of the same colour close together in 1D (either by design, or due to random clustering). 
We now ask what is the minimum density of like-coloured bookmarks needed to form stable domains. To address this question we systematically vary bookmark density and perform simulations with clustered patterns (Fig.~\ref{fig:bookmarks1}B) as these are the most effective way to create domains. Here, $\phi$ varies from $0.01$ to $0.1$ for a chain with $L = 1000$. To facilitate the analysis, we fix the domain size at $100$ beads ($300$ kbp), which is in the range of typical HiC domains~\cite{Dixon2015,Lieberman-Aiden2009,Rao2014}. 
%Then, there can be a maximum of $10$ domains in the simulation (this is achieved by adjusting $n_c$ as $\phi$ changes). 
We set the system to be in the collapsed-ordered phase, i.e. $\epsilon/k_BT= 1$ and $f = 2$, and quantify the efficiency of domain formation by measuring the probability that bead $i$ ($1\le i\le L$) is in a ``red'' state, $P_{\rm red}(i)$. If ideal regular domains are formed along the fibre (i.e., if all beads have the intended colour, that of the closest bookmarks) then $P_{\rm red}(i)$ would be a perfect square wave $\Pi(i)$ (Fig.~\ref{fig:accuracy}, caption).
%\begin{equation}
%\Pi(i) = \dfrac{1}{2} \left[ \sgn{ \left[ \sin{ \left( \dfrac{\pi i}{n_d} \right) } \right] } + 1 \right]
%\end{equation}
%where $n_d$ is the number of beads in a domain (here $n_d=100$). 
The fidelity of domain formation can then be estimated as $\chi=1-\Delta^2$, where $\Delta^2$ is the mean square deviation (variance) between $P_{\rm red}(i)$, measured in simulations, and $\Pi(i)$, i.e. $\Delta^2 = \sum_{i=1}^L \left[P_{\rm red}(i)-\Pi(i)\right]^2/L$.
The fidelity parameter is $\chi \simeq 1/2$, when the epigenetic pattern is far from the ordered block-like state and is dominated by a single colour, whereas $\chi \simeq 1$ for ideal block-like domain formation.

We observe (Fig.~\ref{fig:accuracy}A) that the system displays a phase transition near the critical density $\phi_c \simeq 0.04$. For $\phi > \phi_c$, stable domains are seen in kymographs and $\chi \simeq 1$. For $\phi< \phi_c$ instead, a single mark takes over the whole fibre. Near $\phi=\phi_c=0.04$ there is a sharp transition between these two regimes in which domains appear and disappear throughout the simulation (see kymograph in Fig.~\ref{fig:accuracy}B). 

The critical density $\phi_c$ corresponds to about $1$ or $10$ nucleosomes in about $400$ as not all nucleosomes coarse-grained in a ``bookmark bead'' need to be bookmarked. We argue that, crucially, not all the genome must have this critical density of bookmarks, but only regions required to robustly develop a specific domain of coherent histone marks in a given cell-line.  

%?? What happens if the number of domains is different from 10? With a single bookmark colour, what is the density required to colour red most of the simulations/induce symmetry breaking in practice? %%
%Do we have an issue between what is active and inactive bookmark? Bookmarking regions should be DHS presumably, how can this work for H3K9me3? ?? %% no
%%fig 4 critical bookmark density

%fig 4: heterochrom +bookmarks replication and excision
\begin{figure*}[t!]
	\centering
	\includegraphics[width=0.8\textwidth]{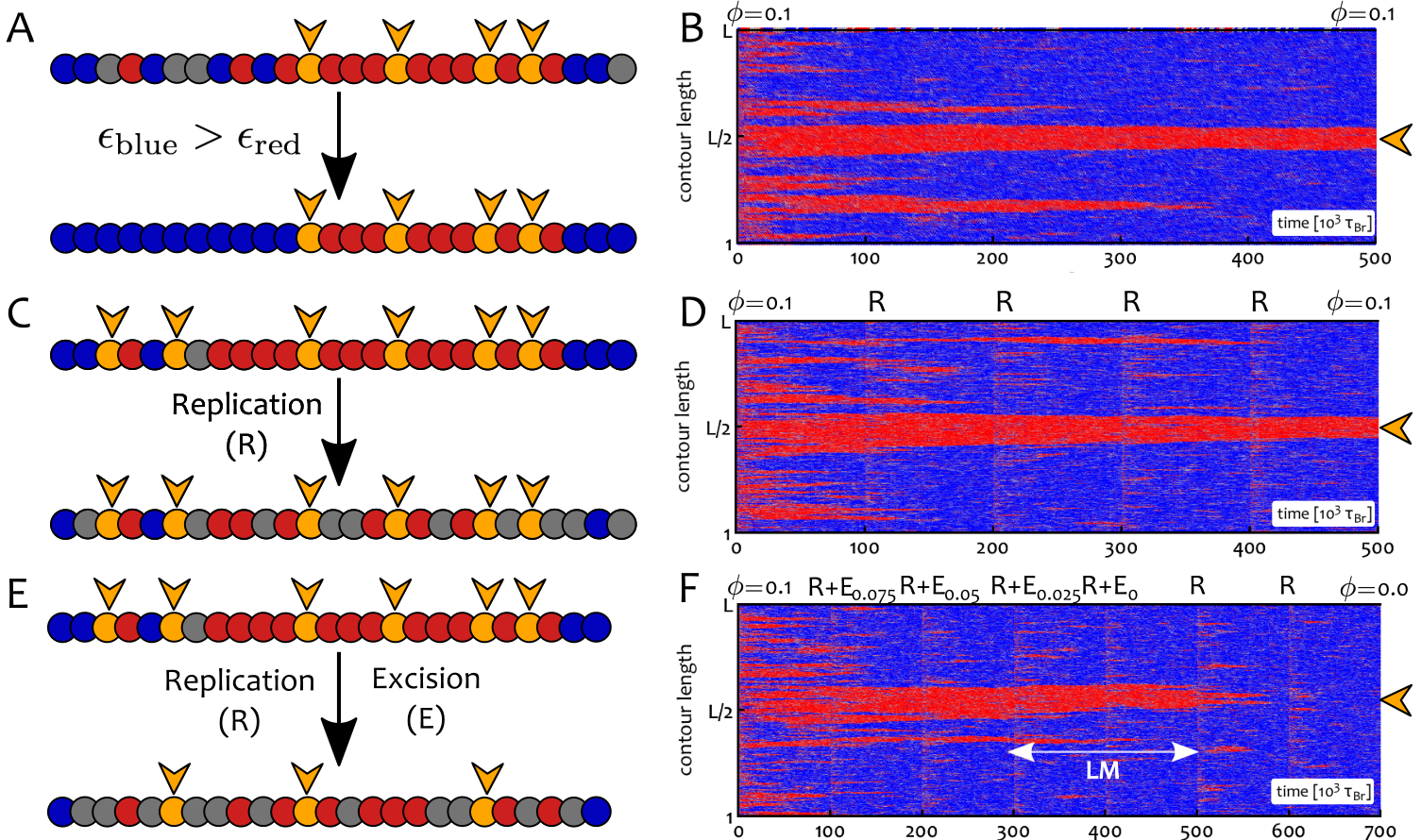}
	\caption{\textbf{Asymmetric Interactions and Bookmark Excision but not DNA Replication Affect the Epigenetic Landscape.} (\textbf{A-B}) Here we consider the case in which blue-blue interactions are stronger than red-red ones. We set $\epsilon_{\rm blue}=1 k_BT$ and $\epsilon_{\rm red}=0.65 k_BT$ with $f=2$. The central region of a chromatin segment $L=2000$ beads long is initially patterned with bookmarks at density $\phi=0.1>\phi_c$ (this region is indicated in the kymograph by an orange arrowhead). Blue beads invade non-bookmarked regions thanks to the thermodynamic bias whereas the local red state is protected by the bookmarks. (\textbf{C-D}) The chromatin fibre undergoes replication cycles which extensively perturb the pattern of PTM of histones on chromatin. A semi-conservative replication event (R) occurs every $10^5$ $\tau_{Br}$ and half of the (non-bookmarked) beads become grey. The epigenetic pattern is robustly inherited. (\textbf{E-F}) The chromatin fibre undergoes semi-conservative replication followed by excision of bookmarks (R+E). At each time, 1/4 of the initial bookmarks are removed and turned into grey (recolourable) beads. The epigenetic pattern is inherited until $\phi<\phi_c$. At this point, the central red domain is either immediately lost (not shown) or it can be sustained through some replication cycles (\textbf{F}) by local memory (LM). \red{See also Suppl. Movie 5 for a direct comparison of the behaviour with and without bookmarks.}}% This entails that bookmarks are crucial element that allow epigenetic memory. In addition, this figure suggests that activated bookmarks may be able to disrupt an homogeneous landscape and establish \emph{de novo} transient epigenetic domains~\cite{Ciabrelli2017}.}
	\label{fig:excision}
\end{figure*}

%%fig 5 Real drosophila chromosome 
\begin{figure*}[t!]
	\centering
	\includegraphics[width=0.83\textwidth]{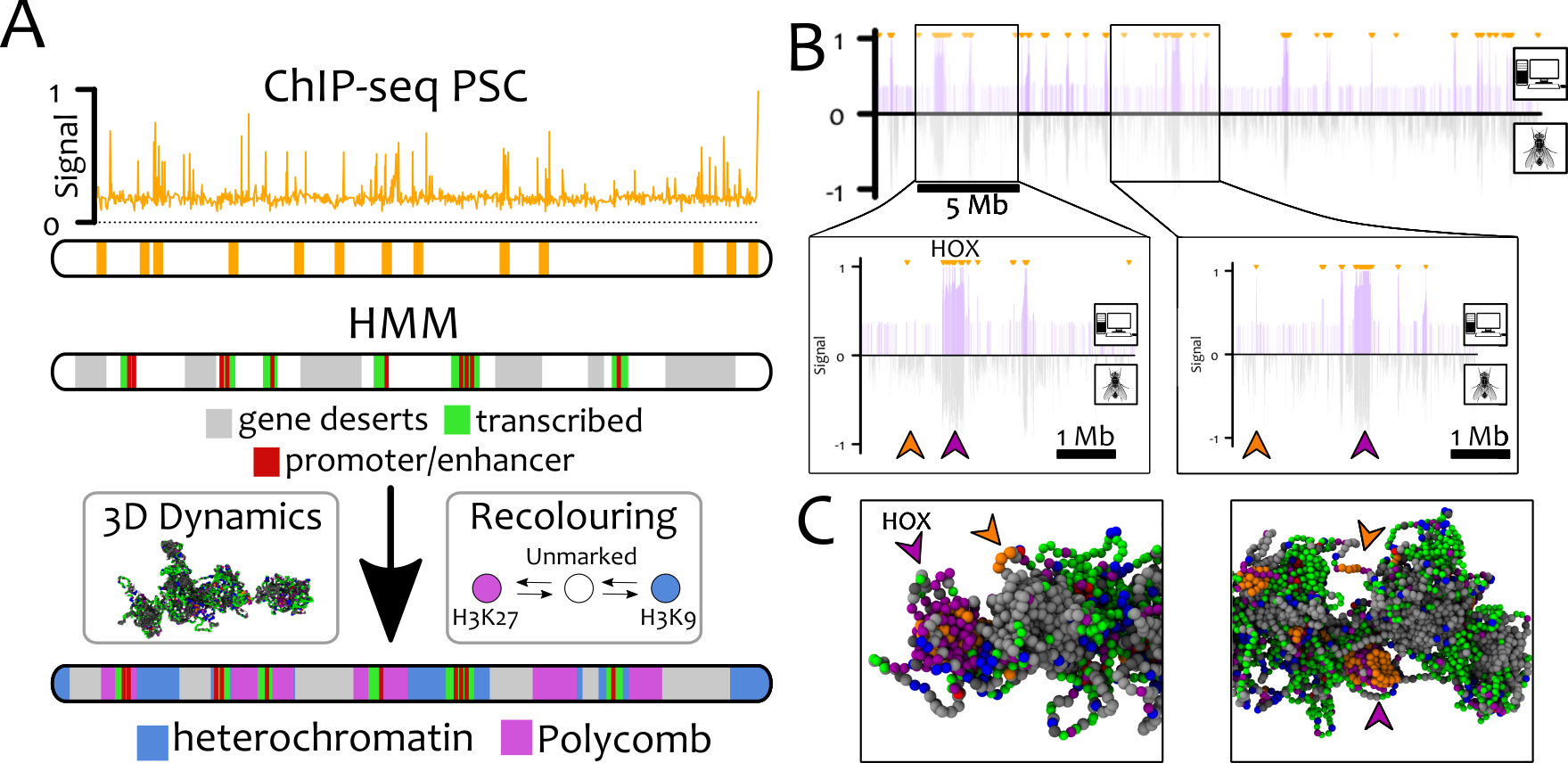}
	\caption{\textbf{GBM Alone is Able to Recapitulate the Distribution of Polycomb Marks in \emph{Drosophila} S2 cells.} Here we perform chromosome-wide simulations of Ch3R of Drosophila S2 cells at $3$ kbp resolution ($L=9302$) with GBM. (\textbf{A}) The location of PSC/PRE (bookmarks) are mapped onto beads using ChIP-Seq data from Ref.~\cite{Follmer2012}. Using the ``9-states'' HMM data~\cite{Kharchenko2011}, gene deserts (regions lacking any mark in ChIP-seq data, state 9), promoter/enhancers (state 1) and transcriptionally active regions (states 2-4) are permanently coloured grey, red and green, respectively. The remaining beads ($\sim$20\%) are initially unmarked (white) and may become either heterochromatin (blue) or polycomb (purple). (\textbf{B}) \emph{In silico} ChIP-seq data for H3K27me3 (top half, purple lines) is compared with \emph{in vivo} ChIP-seq~\cite{Kharchenko2011} (bottom half, grey line). Small orange arrows at the top of the profile indicate the location of the bookmarks. The excellent quantitative agreement between the datasets is captured by the Pearson correlation coefficient $\rho=0.46$ -- to be compared with $\rho=0.006$ obtained between a random and the experimental datasets. We highlight that not all the bookmarked beads foster the nucleation of H3K27me3 domains (see big purple/orange arrowheads in the insets, corresponding to the HOX cluster). The reason can be found by analysing the 3D conformations of the chromosome (\textbf{C}). The non-nucleating bookmarks (orange arrowheads), although near in 1D, are found far from potential target beads in 3D space (purple arrowheads) and so fail to yield large H3K27me3 domains. \red{See also Suppl. Movie 6 for a direct comparison of the results with and without bookmarks.} }
%This suggests an intriguing relationship between 1D information and 3D organisation in preserving epigenetic memory~\cite{Ciabrelli2017}. We finally speculate that the these bookmarks may be functioning as nucleation points in other cell lines which display different 3D organisations.
%The p-value testing independence between the simulated and experimental data-sets is $p=3$ $10^{-13}$, to be compared with $p=0.75$ obtained using a random data-set against the experimental one.
	\label{fig:drosophila}
\end{figure*}

\subsection{Biasing Epigenetic Landscapes with Asymmetric Interactions}

Thus far, we have considered symmetric interactions between like-coloured beads. In other words, red-red and blue-blue interaction strengths were equal. However, such binding energies may differ if mediated by distinct proteins.
Consider the case where red and blue marks encode Polycomb repression and constitutive heterochromatin, respectively. If the blue-blue interaction is larger than the red-red one, the thermodynamic symmetry of the system is broken and the blue mark eventually takes over all non-bookmarked regions (Fig.~\ref{fig:excision}A). However, if there are bookmarks for the red mark, they locally favour the red state, whereas the stronger attraction globally favours the blue mark. This competition creates an additional route to form stable domains as exemplified in Figure~\ref{fig:excision}A,B. Here, red bookmarks (identified by orange beads) are concentrated in the central segment of a chromatin fibre. Starting from a swollen and epigenetically disordered fibre, where red, blue and grey beads are equal in number, we observe that blue marks quickly invade non-bookmarked regions and convert red beads into blue ones (a process mimicking heterochromatic spreading \emph{in vivo}~\cite{Hathaway2012}). However, the central segment containing the bookmarks displays a stable red domain (Fig.~\ref{fig:excision}A,B). 
%Interestingly, bookmarking is necessary but not sufficient to yield stable coexistence between domains, as we further require that the attraction between red bookmarks and red beads is large enough to locally overcome the thermodynamic preference of the blue marks.

\subsection{Bookmark Excision but not DNA Replication Destabilises the Epigenetic Landscape}

We next asked whether the epigenetic pattern established through GBM is also stable against extensive perturbations such as DNA replication. In order to investigate this scenario we simulated semi-conservative replication of the chromatin fibre by replacing half of the (non-bookmarked) beads with new randomly coloured beads~\cite{Berry2017}. In Figure~\ref{fig:excision}C-D we show that our model can ``remember'' the established epigenetic pattern through multiple rounds of cell division. Importantly, the combination of ``memory'' and local epigenetic order (via bookmarks) may allow cells to display ``epialleles'', i.e., alleles with different transcriptional behaviours thus explaining local (or ``cis-'') memory~\cite{Berry2015,Berry2017}. 

We next considered a set-up relevant in light of recent experiments in {\it Drosophila}~\cite{Laprell2017,Coleman2017}, where the role of Polycomb-Response-Elements (PREs) in epigenetic memory was investigated. In these works, polycomb-mediated gene repression was perturbed as a consequence of artificial insertion or deletion of PREs. In Figure~\ref{fig:excision} we thus performed a simulated dynamic experiment where replication was accompanied by random excision of bookmarks~\cite{Laprell2017} (Fig.~\ref{fig:excision}E,F); in practice, we remove $1/4$ of the initial number of bookmarks at each replication event. Then each ``cell cycle'' successively dilutes the bookmarks which at some point can no longer sustain the local red state and the region is consequently flooded with blue marks. 

Importantly, the system does not display immediate loss of the red domain as soon as $\phi<\phi_c$; on the contrary, this domain is temporarily retained through local memory (see Fig.~\ref{fig:excision}F, LM)~\cite{Angel2011,Berry2015,Berry2017}. This originates from an enhanced local density of marks together with the positive read/write feedback (see SM). [The persistence of the local memory can be tuned via the parameters of our polymer model.] These results are again consistent with experiments, as regions of the \emph{Drosophila} genome marked with H3K27me3 are only gradually lost after PRE excision~\cite{Laprell2017}. Similarly, epialleles have been observed to be temporarily remembered across cell division~\cite{Berry2015}. 

\red{We finally highlight that the results presented in Fig.~\ref{fig:excision} are independent on the chosen initial configuration. In SM (Figs.S4-S5) we show that starting from a collapsed and epigenetically disordered chromatin (CD phase), resembling heavily condensed and sparsely marked mitotic structures, leads to the same behaviour and strongly supports the robustness of our findings.}

\subsection{Chromosome-Wide Simulations Predict the Epigenetic Landscape in \emph{Drosophila}}

Simplified models considered thus far are useful to identify generic mechanisms; we now aim to test our model in a realistic scenario. To do so, we perform polymer simulations of the whole right arm of chromosome 3 in {\it Drosophila} S2 cells. 

Bookmarks (orange, in Fig.~\ref{fig:drosophila}) are located on the chromosome using PSC ChIP-Seq data~\cite{Follmer2012}, as PSC binds to PREs during inter-phase and mitosis~\cite{Follmer2012} as well as recruiting PRC2 (via molecular bridging). Some other beads are permanently coloured according to the ``9-state'' Hidden Markov Model (HMM,~\cite{Kharchenko2011}). If they correspond to gene deserts (state 9), promoter/enhancers (state 1) or transcriptionally active regions (states 2-4) they are coloured grey, red and green, respectively. We further introduce an interaction between promoter and enhancer beads to favour looping, plus, an attractive interaction between gene desert (grey) beads mimicking their compaction by H1 linker histone~\cite{Sexton2012} (see SM for full list of parameters). The remaining 20\% of the polymer is left blank and these ``unmarked'' beads are allowed to dynamically change their chromatin state into heterochromatin (blue) or polycomb (purple) according to our recolouring scheme. 

We evolve the system to steady state and we evaluate the probability of finding a Polycomb mark at a certain genomic position. [To determine these probability, a bookmarked bead is counted as bearing the H3K27me3 mark when it is near beads with polycomb marks, or within large stretches of bookmarked beads]. This provides us with an \emph{in silico} ChIP-seq track for Polycomb marks which can be compared with \emph{in vivo} ChIP-Seq data~\cite{Kharchenko2011} (see Fig.~\ref{fig:drosophila}B). The two are in excellent agreement (Pearson correlation coefficient $\rho=0.46$, against $\rho=0.006$ for a random dataset). 

Remarkably, not all bookmarked segments (orange) are populated by Polycomb marks; instead we observe that H3K27me3 spreading requires appropriate 3D folding (Fig.~\ref{fig:drosophila}B-C, insets). Bookmarks which do not contact other bookmarks due to the local epigenetic landscape do not nucleate H3K27me3 spreading. Again, this is consistent with 3D chromatin conformation being crucial for the spreading and establishment of epigenetic patterns~\cite{Ciabrelli2017,Engreitz2013,Deng2014}. 

\section{Discussion}

We proposed and investigated a new biophysical mechanism for the \emph{de novo} establishment of epigenetic domains and their maintenance through interphase and mitosis. Our simplest model requires only one element: a positive feedback between readers (e.g., binding proteins HP1, PRC2, etc.) and writers (e.g., methyltransferases SUV39, EzH2, etc.). 

We performed large-scale simulations in which chromatin is modelled as a semi-flexible bead-and-spring polymer chain overlaid with a further degree of freedom representing a dynamic epigenetic patterning. Specifically, each bead is assigned a colour corresponding to the local instantaneous epigenetic state. Readers are implicitly included by setting an attraction between like-coloured beads~\cite{Brackley2016nar,Barbieri2012}, whereas writers are modelled by performing re-colouring moves according to realistic and out-of-equilibrium rules~\cite{Dodd2007,Dodd2011} (see Fig.~\ref{fig:model}). %Intriguingly, we note that the main qualitative behaviours are retained when detailed balance is restored~\cite{Michieletto2016prx} (see SM).

%and (ii) the presence of ``bookmarks'', i.e. DNA-sequence specific transcription factors which can recruit read/write machineries. 

We find that, if read-write positive feedback is sufficiently strong, a single histone mark can spread over the whole fibre and drives a discontinuous transition to a collapsed-ordered state (see Fig.~\ref{fig:phasediagr}). This state is stable and robust against extensive perturbations such as those occurring during replication~\cite{Talbert2006,Zentner2013,Cavalli2013}, when most histones are removed or displaced~\cite{Alberts2014,Zentner2013,Festuccia2017}. In other words, our model displays ``epigenetic memory''.

The main limitation of this simple model is that epigenetic order in real chromosomes is local, rather than global. Distinct epigenetic domains coexist on a chromosome, thereby forming an ``heterogeneous'' epigenetic pattern. Our main result is that this feature of real chromosomes can be reproduced by our model when we include genomic bookmarking (GBM). 

%here the is a lot of peter -- 
%i add ``dynamically associated'' instead of bound
%Bookmarks are usually considered to be transcription factors that remain (dynamically) associated through mitosis so gene activity can be inherited from one interphase to the next~\cite{Kadauke2013a,Deluz2016,Sarge2005}. %There are many classical examples, including PcG~\cite{Lanzuolo2007,Laprell2017}, GATA~\cite{Kadauke2012}, Esrbb~\cite{Festuccia2016}, PSC~\cite{Follmer2012} and Sox2~\cite{Teves2016}. %This list will probably soon expand as many TFs are now know to remain associated to DNA through mitosis – contrary to what had been thought~\cite{Teves2016}.
Here, we envisage bookmarks which can perform functions typical of many TFs: they recruit read/write machineries, and hence nucleate the spreading of epigenetic marks and the establishment of epigenetic domains. We also assumed that bookmarking TFs are permanently bound to DNA, however our conclusions should hold even for dynamic bookmarks that switch between bound and unbound state~\cite{Brackley2016ephe,Teves2016}. 
  
%fig 6 a model for differentiation w bookmarks+pve feedback
\begin{figure*}[t!]
\centering
\includegraphics[width=0.80\textwidth]{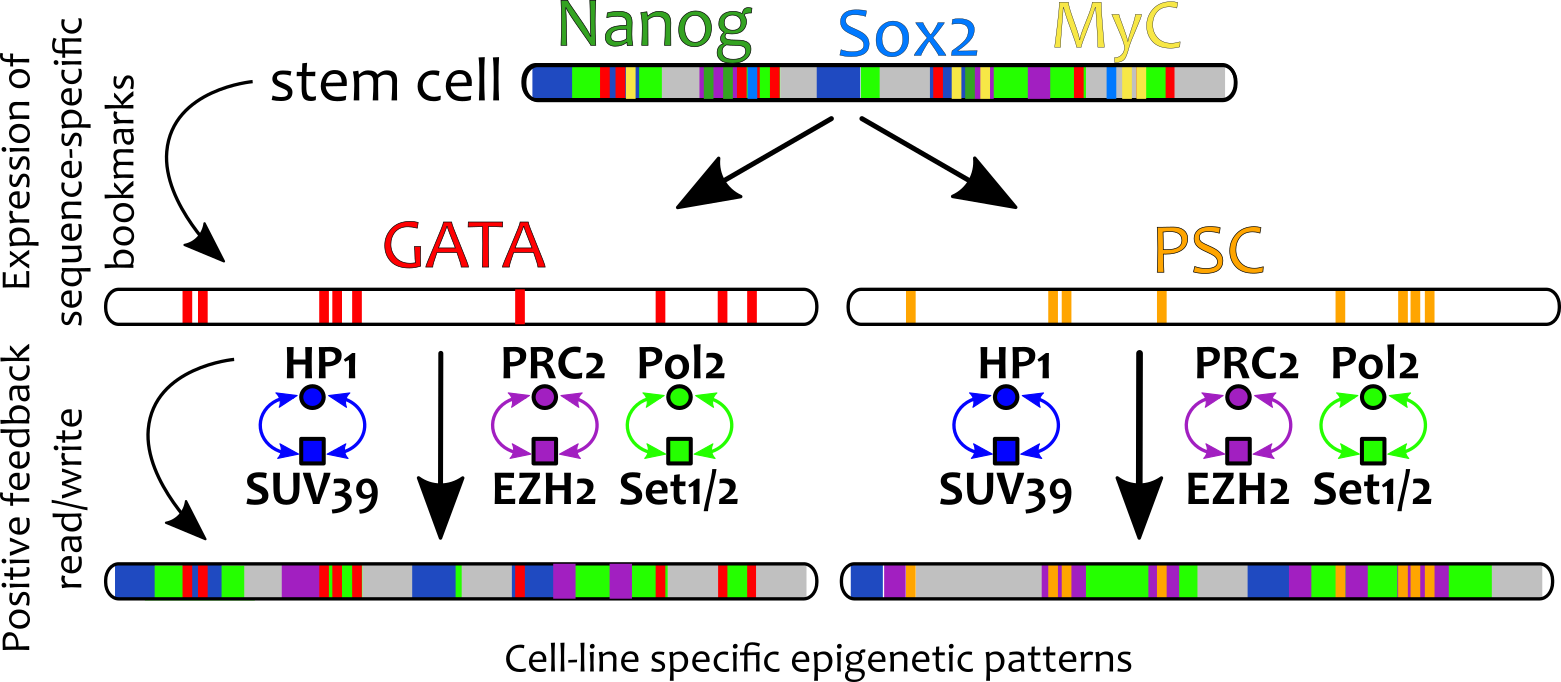}
\caption{\textbf{Model for Cellular Differentiation}. We speculate that cellular differentiation may be driven by a two-step process. First, sequence-specific factors (bookmarks) are expressed as a consequence of environmental and positional cues. Second, the positive feedback set up by read/write machineries drives the establishment and maintenance of tissue-specific epigenetic patterns. As a consequence, genomic bookmarks are key targets to understand cellular differentiation and reprogramming.}
\label{fig:differentiation}
\end{figure*}

%The inclusion of GBM allows us to identify some basic rules required for the establishment of stable epigenetic domains. For instance, GBM must be locally clustered and above a critical density (about one in 400 nucleosomes, Fig.~\ref{fig:accuracy}).  This scenario may be relevant to cases where regions bookmarked with ``repressive'' GBM coexist alongside other regions bookmarked with ``active'' GBM.

We find that stable domains can be formed with only one type of bookmark when the competing epigenetic mark is thermodynamically favoured (Fig.~\ref{fig:excision}). This result rationalises the common understanding that heterochromatin can spread at lengths (blue mark in Fig.~\ref{fig:excision}A,B) and it is stopped by actively transcribed (bookmarked) regions. Further, it is in agreement with recent genome editing experiments in {\it Drosophila}: when PRE is inserted into the genome, it provides a bookmark for H3K27me3 which leads to spreading of that mark~\cite{Laprell2017}, whereas PRE excision leads to (gradual) loss of the mark~\cite{Laprell2017} (Fig.~\ref{fig:excision}). Additionally, the expression of HOX and other Polycomb-regulated genes (which contain multiple PREs) is predicted by our model to be less sensitive to deletion of single PREs~\cite{De2016}. We suggest that this is because domains remain stable if bookmark density is kept above the critical threshold (Fig.~\ref{fig:accuracy}).

%These findings can be readily explained within our model if the 3D attraction between polycomb marks is slightly weaker than that between, e.g., H3K9me3-modified regions (Fig.~\ref{fig:excision}). 

Our results strongly suggest that bookmarks can establish specific epigenetic domains by exploiting the local diffusion of chromatin and thereby ``infecting'' 3D-proximal chromatin segments. The local increase in the density of a mark is then stopped either by thermodynamics (Fig.~\ref{fig:excision}A) or competition with other bookmarks (Fig.~\ref{fig:bookmarks1}B). Crucially, our model does not require any boundary element to stop the spreading of marks, which is instead self-regulated. 

Losing bookmarks (via artificial excision or DNA mutation) will thus impair the ability of cells to inherit the cell-line-specific epigenetic patterns. In addition, we argue that newly activated bookmarks (for instance subsequently to inflammation response or external stimuli~\cite{Kirmes2015,Feuerborn2015,Wood2014}) may drive the \emph{de novo} formation of transient epigenetic domains which allow the plastic epigenetic response to environmental changes. 

We show that our model can recreate the pattern of H3K27me3 in \emph{Drosophila} S2 cells starting solely from the position of PSC proteins acting as Polycomb bookmarks
% transcriptionally-determined pattern of TFs mix of static and dynamic epigenetic marks, with ~20\% of the genome initially being unmarked. Then, our recolouring scheme completes the epigenetic landscape, and quantitatively recapitulates distributions of H3K27me3 seen \emph{in vivo}. 
Intriguingly, our simulations show that not all bookmarks end up in H3K27me3 domains; whether or not they do, depends on their network of chromatin contacts in 3D. This is agreement with recent experiments~\cite{Ciabrelli2017,Engreitz2013,Deng2014} and it is also reminiscent of the well-known position effect according to which the activity of a gene depends on its local environment~\cite{Feuerborn2015}. 

\red{While our framework can be directly applied to model competition between repressive epigenetic marks, the deposition of active marks may be better modelled as resulting from a co-transcriptional positive feedback loop. In light of this, in the SM we show that a model with thermodynamically favoured heterochromatin competing with local recolouring due to transcription leads to results that are qualitatively similar to those presented in the previous sections, as long as promoters are seen as bookmarks for active marks (see SM for more details).}

Our results also prompt several further questions. First, starting from a stem cell, how might different cell lineages be established?  We suggest that environmental and morphological cues trigger production of lineage-specific bookmarks such as GATA~\cite{Kadauke2013a} and PSC~\cite{Follmer2012}, which nucleate the positive feedback between readers and writers to generate and sustain new cell-line specific epigenetic patterns (Fig.~\ref{fig:differentiation}). Thus, bookmarks are here envisaged as key elements that should be targeted in order to understand, and manipulate, cellular differentiation.
Second, how might reprogramming factors like Nanog work? We argue that their binding can ``mask'' the action of pre-existing bookmarks, thereby allowing the establishment of new epigenetic patterns~\cite{Festuccia2016} (see also BioRxiv: https://doi.org/10.1101/127522).

In conclusion, we have extended the existing notion of GBM to include the ability of nucleating the spreading of epigenetic marks by triggering \emph{local} read/write feedback loops. This model predicts the \emph{de novo} establishment of heterogeneous epigenetic patterns which can be remembered across replication and can adapt in response to GBM-targeted perturbations. 

Within our framework, architectural elements such as CTCF~\cite{Alberts2014}, Cohesins~\cite{Fudenberg2016} and SAF-A~\cite{Nozawa2017} may provide the initial 3D chromatin conformation upon which the GBM-driven establishment of epigenetic landscape takes place.

\section{ACKNOWLEDGEMENTS}

We acknowledge the European Research Council for funding (Consolidator Grant THREEDCELLPHYSICS, Ref. 648050). Work in the Papantonis lab is supported by CMMC core funding. The authors thank C. A. Brackley, A. Buckle, N. Gilbert, J. Allan and G. Cavalli for insightful remarks on the manuscript.

\bibliographystyle{nar}
\bibliography{Epigenetics}

\end{document}